\documentclass[twocolumn,showpacs,superscriptaddress,amsmath,amssymb]{revtex4}
\usepackage{graphicx}
\usepackage{dcolumn}
\usepackage{bm}
\usepackage{color}
\usepackage[colorlinks,
           bookmarks=true,
           linkcolor=blue,
           urlcolor=blue,
           anchorcolor=black,
           citecolor=blue]{hyperref}
\usepackage[all]{hypcap}
\usepackage{subfigure}

\begin{document}
\title{Effects of Nuclear Potential on the Cumulants of Net-Proton and Net-Baryon Multiplicity Distributions in Au+Au Collisions at $\sqrt{s_{\text{NN}}} = 5\,\text{GeV}$}

\author{Shu He}
\affiliation{ Key Laboratory of Quark \& Lepton Physics (MOE) and Institute of Particle Physics, Central China Normal University, Wuhan 430079, China}
\author{Xiaofeng Luo}
\email{xfluo@mail.ccnu.edu.cn}
\affiliation{ Key Laboratory of Quark \& Lepton Physics (MOE) and Institute of Particle Physics, Central China Normal University, Wuhan 430079, China}
\author{Yasushi Nara}
\affiliation{Akita International University, Yuwa, Akita-city 010-1292, Japan}
\author{ShinIchi Esumi}
\affiliation{Center for Integrated Research in Fundamental Science and Engineering, \\University of Tsukuba, Tsukuba, Ibaraki 305, Japan}
\author{Nu Xu}
\affiliation{ Key Laboratory of Quark \& Lepton Physics (MOE) and Institute of Particle Physics, Central China Normal University, Wuhan 430079, China}
\affiliation{Lawrence Berkeley National Laboratory, Berkeley, CA 94720, USA}

\begin{abstract}
We analyze the rapidity and transverse momentum dependence for the cumulants of the net-proton and net-baryon distributions in Au+Au collisions at $\sqrt{s_{\text{NN}}} = 5\,\text{GeV}$ with a microscopic hadronic transport (JAM) model. To study the effects of mean field potential and softening of equation of state (EoS) on the fluctuations of net-proton (baryon) in heavy-ion collisions, the calculations are performed with two different modes. The softening of EoS is realized in the model by implementing the attractive orbit in the two-body scattering to introduce a reduction pressure of the system. By comparing the results from the two modes with the results from default cascade, we find the mean field potential and softening of EoS have strong impacts on the rapidity distributions ($\text{d}N/\text{d}y$) and the shape of the net-proton (baryon) multiplicity distributions. The net-proton (baryon) cumulants and their ratios calculated from all of the three modes are with similar trends and show significant suppression with respect to unity, which can be explained by the presence of baryon number conservations. It indicates that the effects of mean field potential and softening of EoS might be not the ingredients that are responsible to the observed strong enhancement in the most central Au+Au collisions at 7.7 GeV measured by the STAR experiment at RHIC. 
\end{abstract}
\maketitle

\section{Introduction}
One of the main interests of relativistic heavy ion collision experiments is to explore the phase structure of the QCD matters. The QCD phase structure can be displayed in the two dimensional QCD phase diagram, in which the temperature $T$ is plotted as a function of the baryon chemical potential $\mu_\text{B}$. Lattice QCD calculations  show that the transition from Quark-Gluon Plasma (QGP) to hadronic phase is a crossover transition~\cite{crossover} when $\mu_\text{B}=0$, and the QCD based model
calculations predict the transition at large $\mu_\text{B}$ is of the first order~\cite{firstorder}. Thus, the crossover region and first order region should be connected by a so called QCD critical point~\cite{QCP_Prediction,science,location}. Due to sign problem in Lattice QCD calculations at finite baryon density, there are large uncertainties to determine the location of critical point by theoretical and/or QCD based model calculations~\cite{location,qcp_Rajiv}. The fluctuation of conserved quantities, such as net-baryon number, and its proxy observable net-proton number, served as the observable sensitive to the correlation length of nuclear matter~\cite{ejiri2006hadronic,qcp_signal,Neg_Kurtosis,Asakawa}, have been extensively studied experimentally~\cite{2010_NetP_PRL,netcharge_PRL,STAR_BES_PRL} and theoretically~\cite{HRG_Karsch,PBM_netpdis,Lattice,2015_JianDeng_fluctuation,2014_Bengt_flu,Asakawa_formula,BFriman_EPJC,2015_Swagato_evolution,2015_Vovchenko,baseline_PRC,huichao,HRG_Nahrgang,kenji_morita,freezeout}. 
\begin{figure}[!htb]
    \hspace{-0.5cm}
	\includegraphics[width=2.4in]{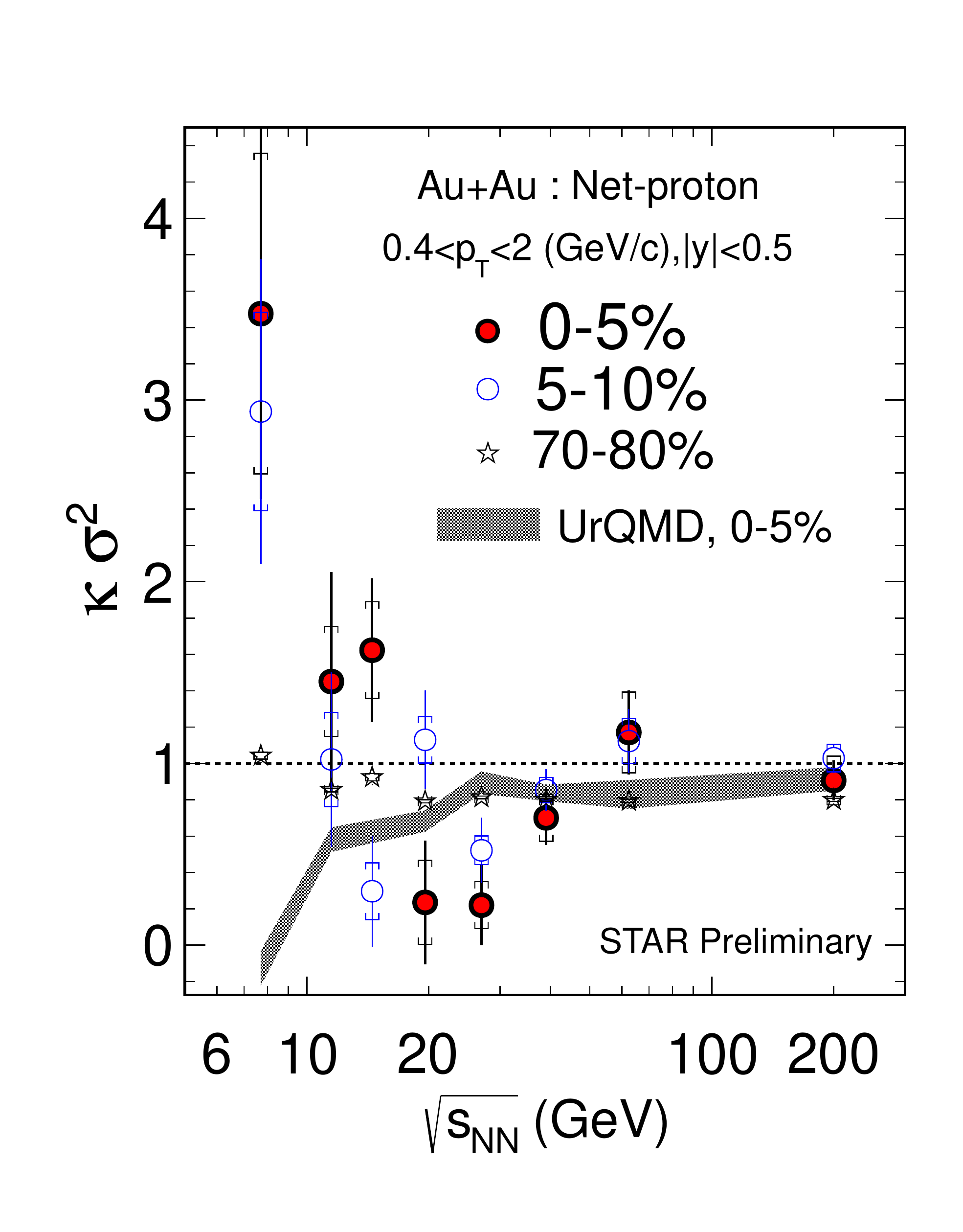}
	\vspace{-0.5cm}
	\caption{Energy dependence of cumulants ratios ($\kappa \sigma^{2}$) of net-proton distributions from RHIC beam energy scan measured by STAR experiment. The dashed shadow band represents the results from UrQMD calculations. Figure is obtained from~\cite{luo2015exploring}.}
	\label{fig:kvsd}
\end{figure}

In recent experimental search for the critical point by the STAR experiment at RHIC, the forth order net-proton fluctuations ($\kappa \sigma^{2}$) displayed a non-monotonic behavior in the energy dependence~\cite{2014_Luo_CPOD,luo2015exploring}. One of the most striking observation is the large deviation above unity of the net-proton $\kappa \sigma^{2}$ in the most central Au+Au collisions at $\sqrt{s_{\text{NN}}} = 7.7\,\text{GeV}$, which is consistent with receiving strong positive critical contribution predicted by theoretical calculations. This observation cannot be described by the UrQMD and AMPT model~\cite{kurtosis_nahrgang,luo2014baseline,urqmd,Jan_kurtosis,UrQMDMoments}, both of which are transport model without mean field and QCD phase transition. This motivates us to investigate whether the large increase in the forth order net-proton fluctuations ( $\kappa \sigma^{2}$) are caused by non-critical contributions, such as mean field potentials, which may play an important role at low collision energies and high baryon density region.  Theoretical model estimated that the size of the correlation length near the critical point is about 2$\sim$3 fm~\cite{2to3fm} in high energy heavy-ion collisions.  One should note that due to the collisions and freeze-out dynamics, even the system indeed passes through the critical region, the critical behavior may not survive in the final observables. It is therefore important to discuss non-critical behavior in order to understand the experimental observations.  On the other hand, it is also important to understand the possible experimental effects such as the efficiency correction and associated systematic errors~\cite{volker_eff}, which are still under studied and not discussed in this paper. 

To study the effects of the mean field potential and softening of equation of state (EoS), we utilized a hadronic transport (JAM) model~\cite{nara1999relativistic} to simulate the net-proton (baryon) fluctuations in Au+Au collisions at $\sqrt{s_{\text{NN}}} = 5\,\text{GeV}$, which is dedicated center of mass energy covered by the future fixed target heavy-ion collisions program in the Compress Baryonic Matter (CBM) experiment~\cite{CBM_experiment} at GSI, Germany and J-PARC experiment~\cite{JPARC} at JAEA/KEK, Japan. The mean field potential and the effects of softening of EoS have been added in Monte Carlo simulation. This studies will allow us to understand the contribution from these two effects to the fluctuations of net-proton (baryon) in heavy-ion collisions. It will provide a baseline and reveal the non-critical background contribution for the baryon number fluctuations when searching for the QCD critical point in heavy-ion collisions. 

The paper is organized as follows. We provide the definition of the cumulant observables in the section II. In the section III, we briefly discuss the JAM model with the mean field potential and the attractive scattering orbit, which is used to simulate the effects of the softening of EoS. Then, the results and discussion will be given in the section IV. Finally, we will summarize our studies in the section V.

\section{Observables}
Fluctuations of net-proton (baryon) can be characterized by their cumulants of the event-by-event multiplicity distribution. It can be computed as
\begin{align}
	C_1 &= \left<N\right> \\
	C_2 &= \left<(\delta N)^2\right> \\
	C_3 &= \left<(\delta N)^3\right> \\
	C_4 &= \left<(\delta N)^4\right> - 3\left<(\delta N)^2\right>^2 
\end{align}
Where $N$ is net-proton (baryon) number and the $\left<N\right>$  is average over events, $\delta N = N-\left<N\right>$. In the STAR experiments, the fluctuations of net-proton number is usually used as a proxy of net-baryon number fluctuations, for the reason that neutral baryons, like neutrons are invisible to the detector. With the definition of cumulants, we can also define mean ($M$), variance ($\sigma^{2}$), skewness ($S$) and kurtosis ($\kappa$) as:
\begin{eqnarray}
M = C_{1}, \sigma^2 = C_{2}, S=\frac{C_{3}}{(C_{2})^\frac{3}{2}}, \kappa = \frac{C_{4}}{(C_{2})^2}
\end{eqnarray}
In addition, the moments product $\kappa\sigma^2$ and $S\sigma$ can be expressed in terms of the ratios of cumulants:
\begin{eqnarray}
\label{eq6} \kappa\sigma^2 = \frac{C_{4}}{C_{2}}, S\sigma = \frac{C_{3}}{C_{2}}, \sigma^{2}/M=\frac{C_{2}}{C_{1}}
\end{eqnarray}
The ratios of cumulants are intensive variables, which is independent on system volume.  The statistical errors of those cumulants and cumulants ratios are evaluated 
by the Delta theorem~\cite{Delta_theory,Unified_Errors}. In general, the statistical uncertainties strongly depend on the shape of the distributions, especially the width of the distributions. For e.g.,  
the statistical errors of various order cumulants ($C_{n}$) can be approximated as $error(C_{n}) \propto \sigma^{n}/(\sqrt{N} \epsilon^{n})$, where $\sigma$ is the measured width of the distribution, $N$ represents the number of events and $\epsilon$ is 
the particle detection efficiency.

\section{JAM Model}
The JAM (Jet AA Microscopic Transportation Model)~\cite{nara1999relativistic} is a hadronic transport model providing a useful tool to study the heavy ion collision from 100A MeV to RHIC energies. The $hh$ inelastic collision part is modeled by hadronic resonance productions at low energies and above this resonance region, the string excitation is adapted. The description of interacting $N$-body system of JAM comes from Relativistic QMD (RQMD) approach ~\cite{isse2005mean, maruyama1995relativistic, maruyama1996relativistic}, which regarded as a hadronic transport model. A simplified version of RQMD (RQMD/S) that focuses in effective using of CPU was adapted in JAM. The Hamiltonian of this $N$-body system has been derived to be
\begin{equation}
H \approx \sum_{i=1}^N u_i(p_i^2 - m_i^2 -2m_i V_i)
\end{equation}
where $u_i=1/2p_i^0$, $p_i^0 = \sqrt{\boldsymbol{p}_i^2 + m_i^2 +2m_i V_i}$. In the mean field case, a Skyrme type density dependent and Lorentzian-type momentum dependent potential was introduced. The total potential energy of the system has the form
\begin{align}
	V &= 
		\int \mathrm{d}\boldsymbol{r}\,
		\left[
		\alpha\frac{\rho^2(\boldsymbol{r})}{2\rho_0}
		+ 
		\beta \frac{\rho^{\gamma+1}(\boldsymbol{r})}{(1+\gamma)\rho_0^\gamma}
		\right] 
		\nonumber \\
	&+ 
		\sum_{k=1,2}{\frac{C_k}{2\rho_0}}
		\int \mathrm{d}\boldsymbol{r}\,\mathrm{d}\boldsymbol{p}\,\mathrm{d}\boldsymbol{p}'
		\frac{f(\boldsymbol{r},\boldsymbol{p})f(\boldsymbol{r},\boldsymbol{p}')}{1+[(\boldsymbol{p}-\boldsymbol{p}')/\mu_k]^2}
\end{align}
where $\rho(\boldsymbol{r})$ is the baryon density distribution. $\alpha$, $\beta$, $\gamma$, $\rho_0$, ${C_k}$, $\mu_k$ are empirical parameters same as reference \cite{nara2015does}.

The effect of the softening of EoS has been investigated in the paper \cite{nara2016directed}. Accompanying with the occurrence of EoS softening, the system usually experienced a first order phase transition\cite{rischke1995phase}. In the JAM model, extra pressure of the system in addition to the free streaming is obtained by two-body scattering given by the virial theorem \cite{nara2016directed}. The attractive orbit in two-body scattering is introduced in the JAM model to simulate the effect of the softening of EoS. By doing this, the pressure of the system can be reduced and the negative slope of rapidity dependence of directed flow are successfully reproduced.

In this work, data of Au+Au collisions at $\sqrt{s_{\text{NN}}} = 5 \,\text{GeV}$ were generated with JAM model in the three modes, which are cascades, mean field potential and attractive scattering orbit, respectively. 

\begin{figure}[htb]
\includegraphics[height=2.8in]{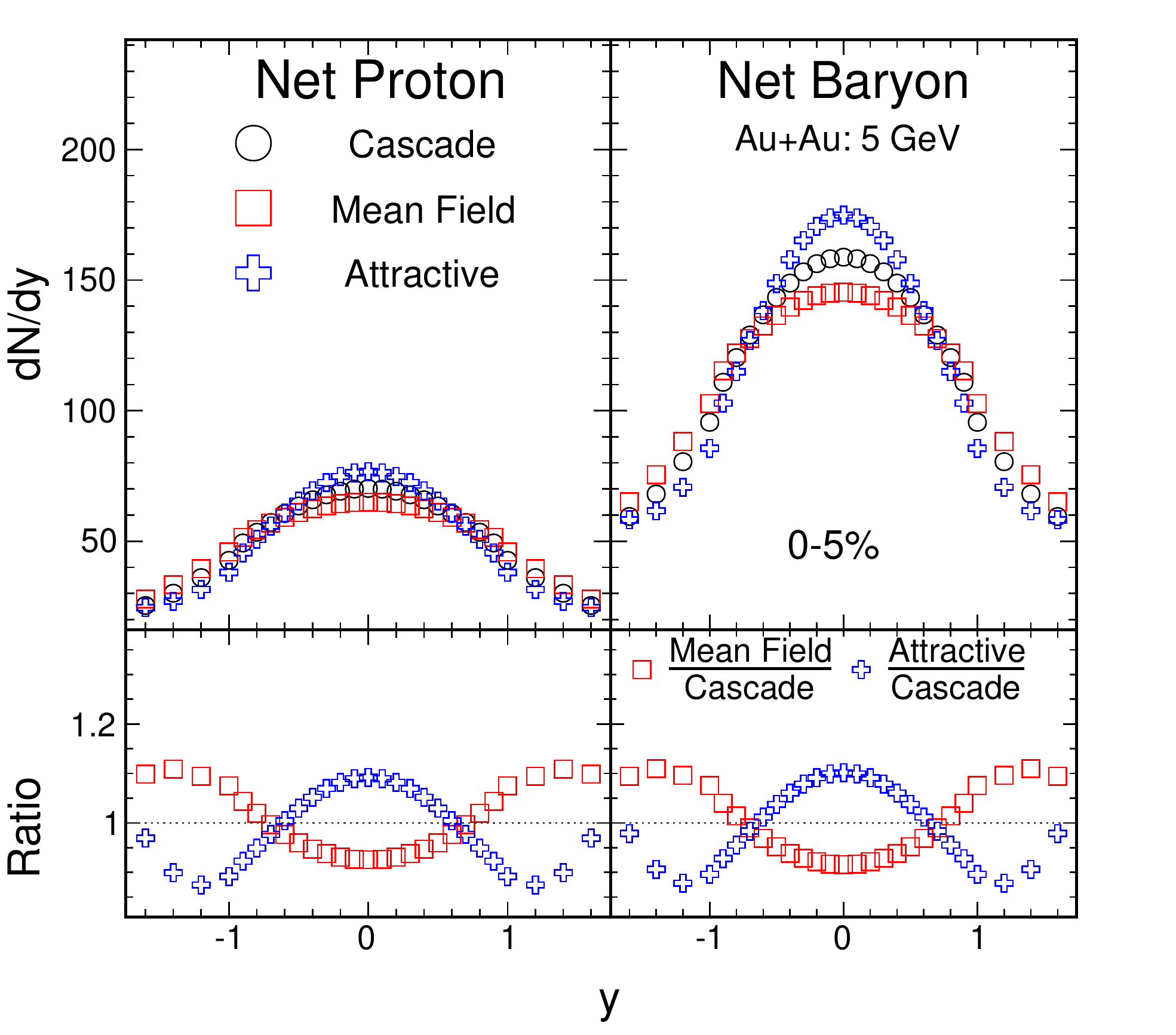}	
\includegraphics[height=3in]{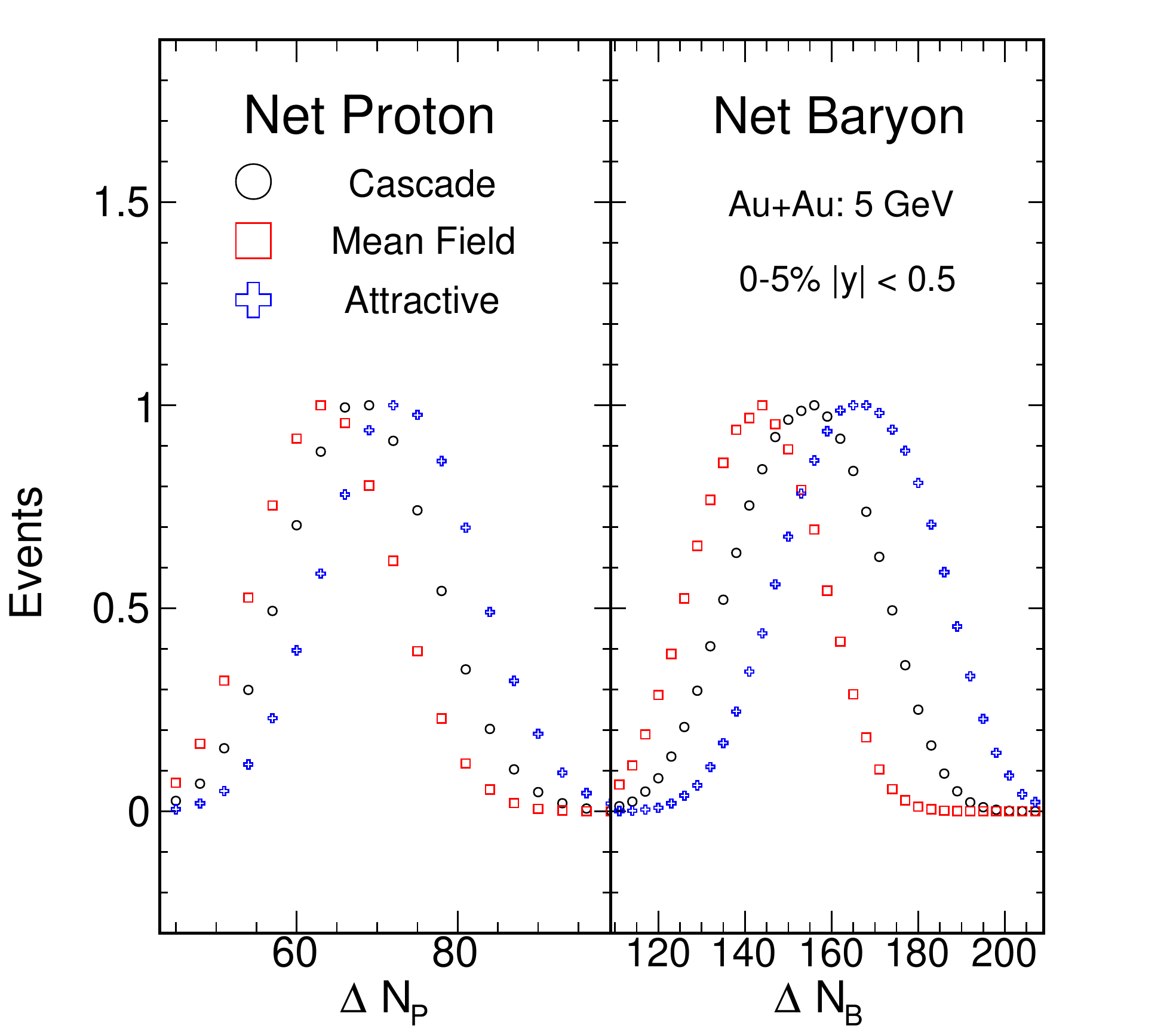}	
	\caption{Rapidity distribution (top) of net-proton (baryon), and event-by-event net-proton (baryon) distributions (bottom) from JAM model simulation for three different modes.}
	\label{fig:fdndy}
\end{figure}

\section{Results}
The calculation results of net-proton (baryon) cumulants would be modified by including the mean field potential in cascade model. This additional mean field potential might have significant effects on the fluctuations of particle multiplicity at low collision energies, and the rapidity distribution of particles in final state. Since the event-by-event net-proton (baryon) cumulants are calculated within certain rapidity window or transverse momentum cuts in order to archive the condition of grand canonical ensemble. It is reasonable to postulate that the calculation results may depend on the types of potentials that applied to the Monte Carlo simulation.
\begin{figure*}[htb]
	\hspace{-0.7cm}
		\includegraphics[height=3. in]{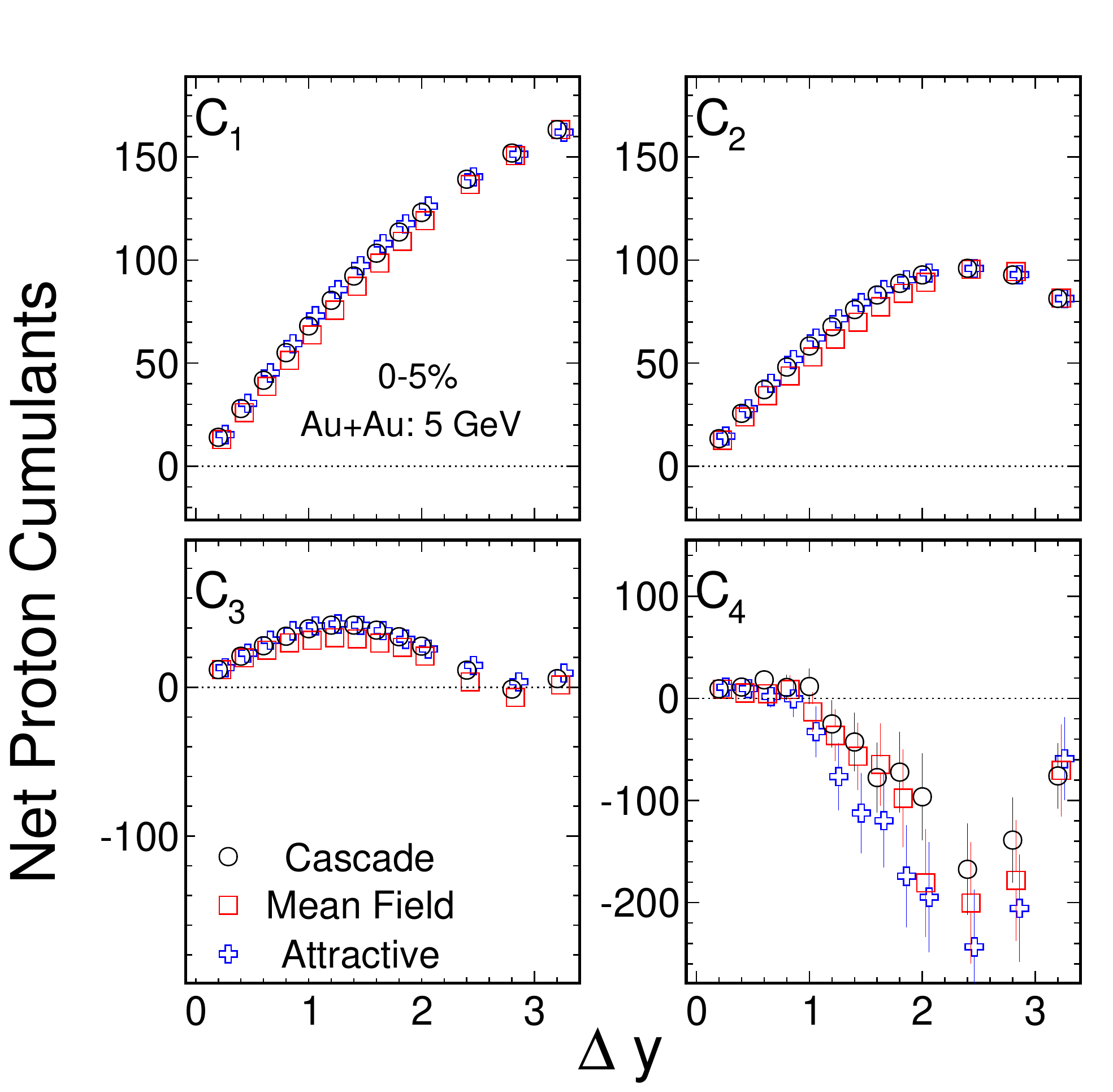}
		\hspace{1.2cm}
		\includegraphics[height=3.in]{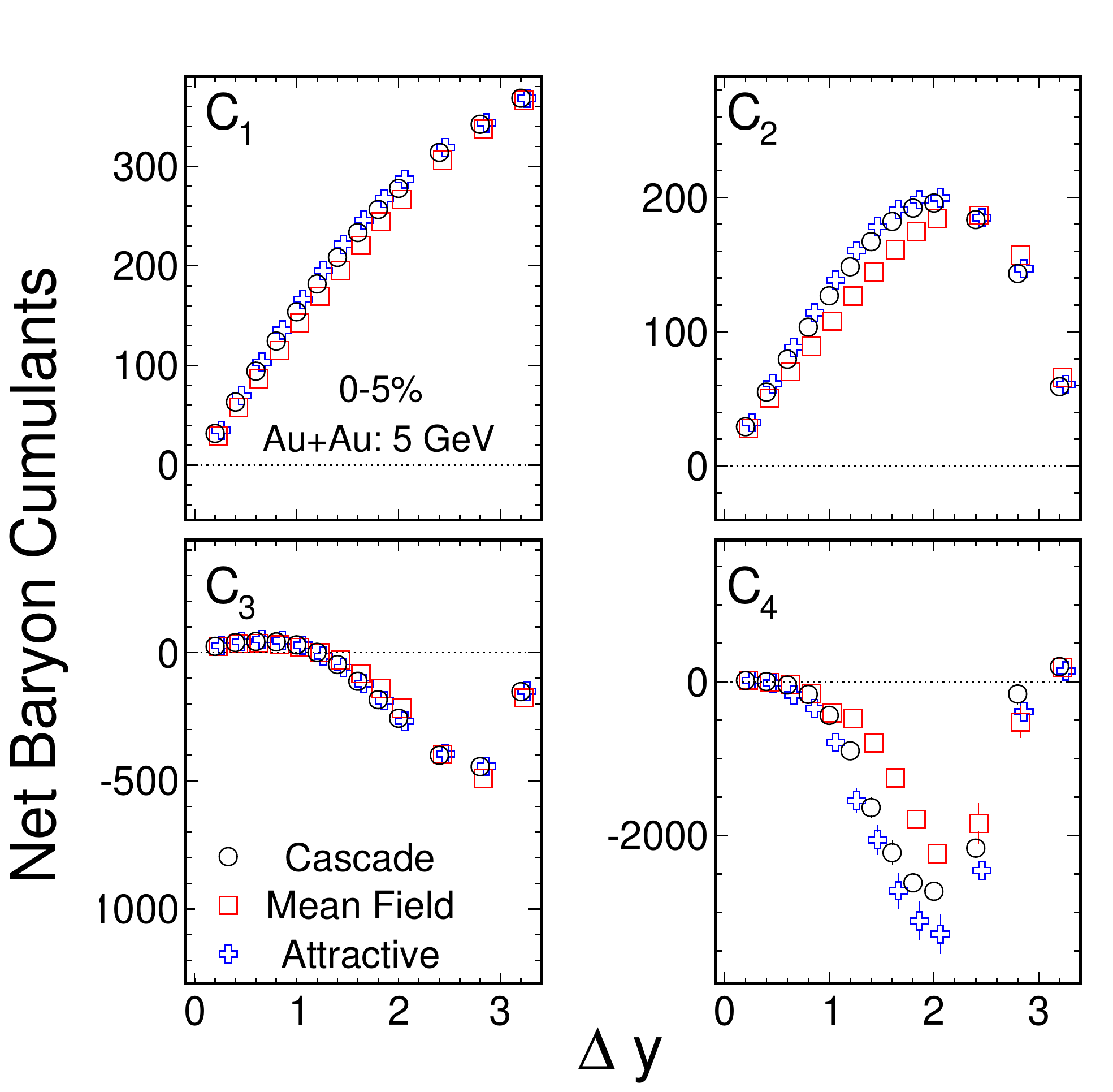}
		\caption{Rapidity dependence for the cumulants ($C_1 \sim C_4$) of net-proton (left) and net-baryon (right) distributions in the most central Au+Au collisions from JAM model with cascade, mean field potential and attractive scattering orbit, respectively.
		The dashed horizontal lines are with the value of zero. }
	\label{fig:cul2y}
\end{figure*}

\begin{figure*}[htb]
		\includegraphics[height=2.6in]{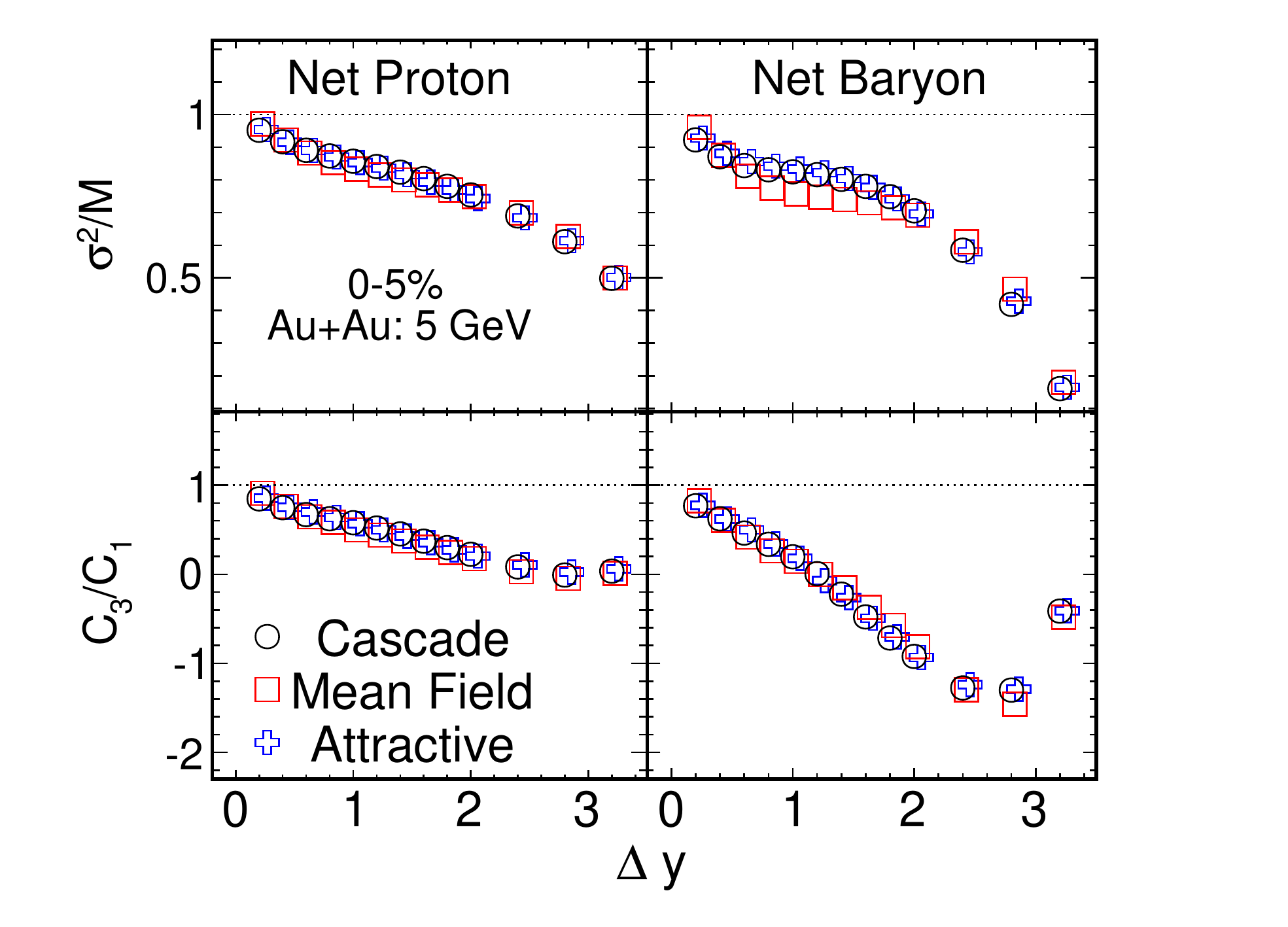}
		\hspace{-1.cm}
		\includegraphics[height=2.6in]{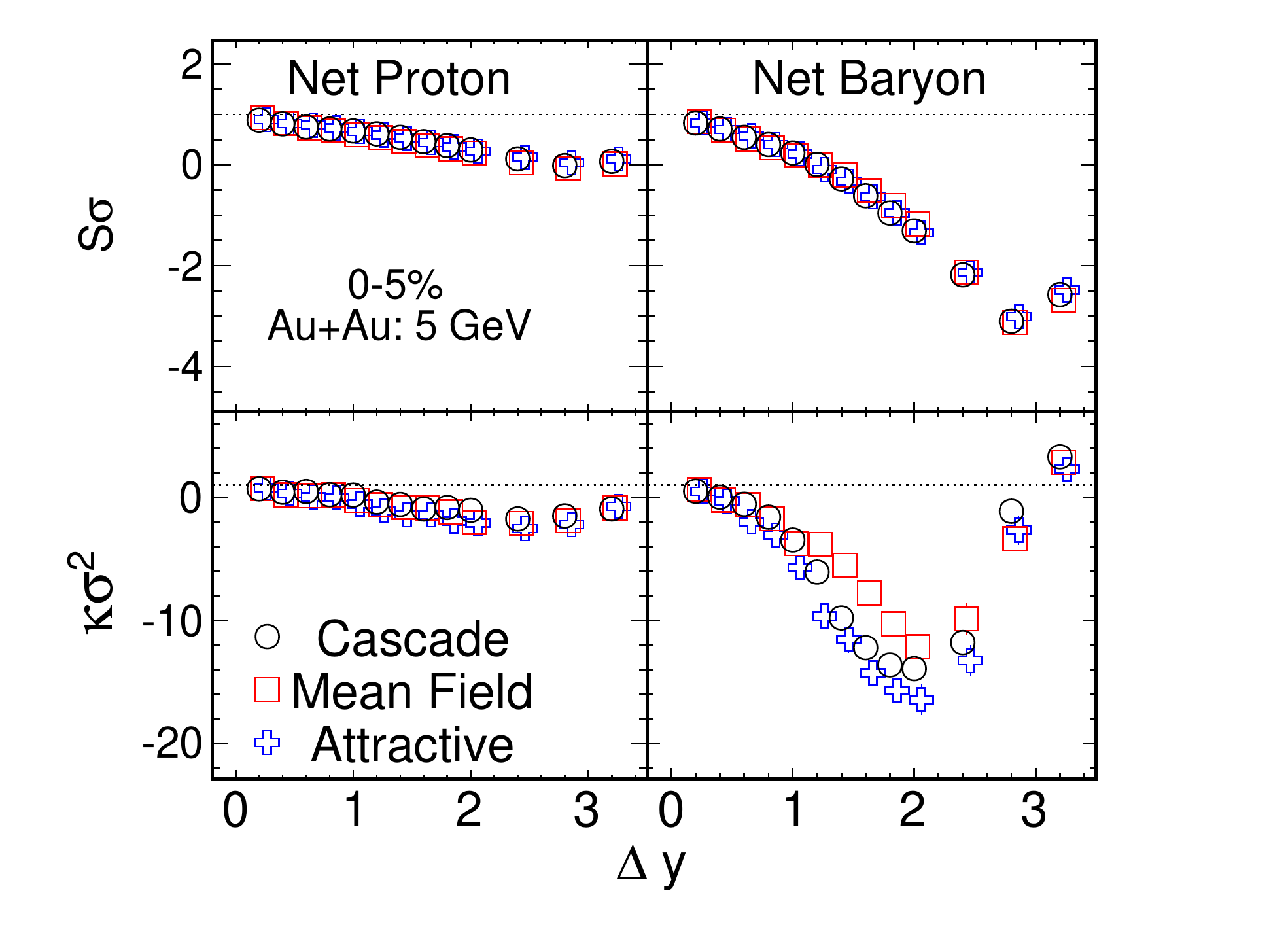}
	\caption{Rapidity dependence for the Cumulants ratios of net proton and net baryon multiplicity distributions in Au+Au collisions at $\sqrt{s_{\text{NN}}} = 5\,\text{GeV}$ GeV from JAM model computed in the three different modes. In the left shows $\sigma^{2}/M$ ($C_2/C_1$) and $C_3/C_1$. The figure in the right shows $S\sigma$ ($C_3/C_2$) and $\kappa\sigma^2$($C_4/C_2$) . The dashed horizontal lines are with the value of unity.} \label{fig:ratios}
\end{figure*}
To avoid the effects of auto-correlation, the collision centralities in the simulation are determined by the multiplicities of charged pion and kaon within pseudo-rapidity $|\eta|<1$ (refmult3) in Au+Au collisions, which is also applied for the data analysis of the net-proton fluctuation in the STAR experiment~\cite{STAR_BES_PRL}. In our study, we will focus on discussing the results from the 0-5\% most central Au+Au collisions, which is defined by the top 5\% fraction of the refmult3 multiplicity distribution for the minimum-biased event ensemble. The volume fluctuations are suppressed by applying the so called centrality bin width correction~\cite{technique}. In the top of the figure~\ref{fig:fdndy},  we show the rapidity distributions ($\text{d}N/\text{d}y$) for the net-proton (baryon) of the most central (0-5\%) Au+Au collisions from JAM cascade, mean field potential and attractive scattering orbit, respectively. The ratios between cascade and the other two cases are displayed in lower panels, respectively. It shows that both the mean field potentials and attractive scattering orbit have strong impacts on the $\text{d}N/\text{d}y$ distributions for net-proton (baryon). Due to the nucleon repulsive potentials, the results from mean field yield less stopping thus wider $\text{d}N/\text{d}y$ distributions than cascade. It is observed that the magnitude of the $\text{d}N/\text{d}y$ from mean field is smaller than the cascade around mid-rapidity ($|y|<0.5$). On the other hand, the attractive orbit scattering results in significant enhancement  with respect to the cascade at mid-rapidity and a narrower $\text{d}N/\text{d}y$ distributions. The enhancement of the $\text{d}N/\text{d}y$ at mid-rapidity for attractive orbit can be attributed to the reduction of the pressure of the system and thus stronger nucleon stopping. We also have studied the event-by-event net-proton (baryon) distributions at mid-rapidity for central Au+Au collisions with JAM model in the three modes. The net-proton (baryon) distributions show similar shape for cascade and mean field cases, while the net-proton (baryon) distributions calculated from attractive scattering orbit case show larger mean values comparing with the distributions from cascade and mean field data.

\begin{figure*}[htb]
\hspace{-0.5cm}
		\includegraphics[height=4.3in]{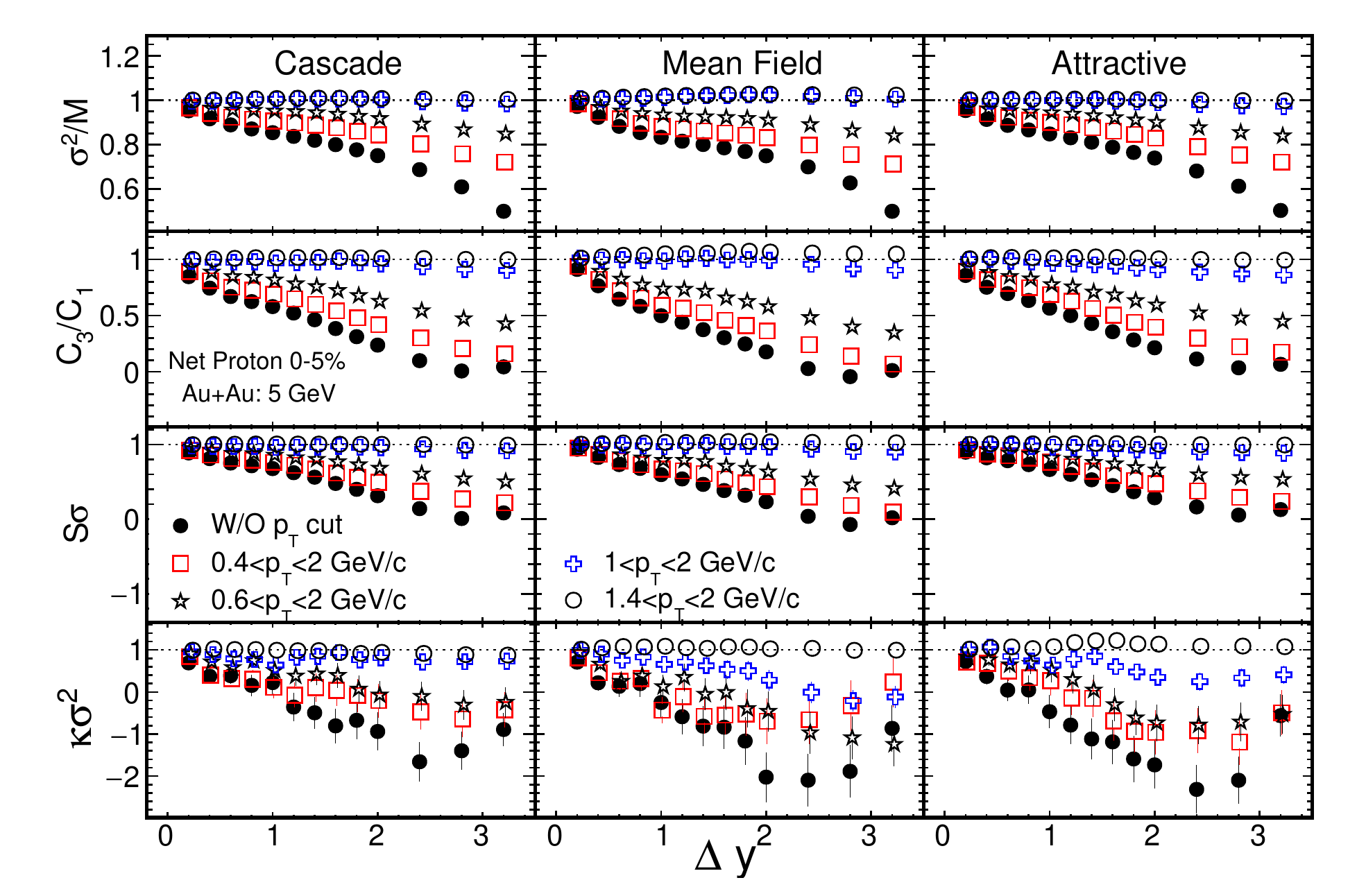}
	\caption{Rapidity dependence for the cumulants ratios of net-proton distributions in Au+Au collisions at $\sqrt{s_{NN}}=5$ GeV from JAM model computed in the three different modes and various transverse momentum ranges. From top to bottom are  $\sigma^{2}/M$ ($C_2/C_1$),  $C_3/C_1$, $S\sigma$ ($C_3/C_2$) and $\kappa\sigma^2$($C_4/C_2$), respectively. The dashed horizontal lines are with the value of unity.}
	\label{fig:pt}
\end{figure*}
Figure~\ref{fig:cul2y} displays rapidity dependence for the cumulants (up to forth order) of net-proton and net-baryon distributions from JAM model computed with three different modes. The rapidity coverage $\Delta y$ of particles are centered at zero and the rapidity cut in the analysis is $|y|< \Delta y/2$. It is found that the net-proton (baryon) cumulants for the three modes show similar trends. Since cumulants are sensitive to number of particles in analyze, the net-baryon cumulants show larger values than that of net-protons. With increasing the rapidity acceptance $\Delta y$, cumulants are rising linearly, for the reason that they are proportional to the volume of the system. The further increasing on $\Delta y$ will lead to significant suppressions due to the effects of baryon number conservation. 

Figures~\ref{fig:ratios} shows cumulant ratios of net-proton (baryon) distributions in Au+Au collisions at $\sqrt{s_{\text{NN}}} = 5 \,\text{GeV}$ from JAM model. Those ratios of cumulants of net-proton (baryon) distributions are constructed to eliminate the volume dependence and can be used to compare with the theoretical calculations. When increasing the rapidity acceptance ($\Delta y$), the net-proton (baryon) cumulant ratios will decrease,  reach a minimum and then increase, which is the typical effects of baryon number conservation~\cite{bzdak2013baryon}. For different net-proton (baryon) cumulant ratios, the position of the minimum are different. It indicates the mean field potential and softening of EoS will not lead to large increase above unity for the net-proton (baryon) cumulants ratios.  Instead, due to the baryon number conservation, large suppression for the fluctuations of net-proton (baryon) are observed. The rapidity dependence for the cumulants ratios calculated from the three modes are with the similar trend. It suggests that the observed similar trends obtained by JAM model without implementing critical physics are dominated by the effects of baryon number conservation. On the other hand, one observes that the net-baryon cumulant ratios show larger suppression with respect to unity than the net-proton and the higher order cumulant ratios also show larger suppression than the lower order. Since the mean field potential implemented in the JAM model is momentum dependent, it is important to study the momentum dependence for the cumulants of net-proton distributions. In Fig. \ref{fig:pt},  for different transverse momentum range, we plot the cumulant ratios of net-proton distributions as a function of rapidity coverage, which are calculated with the three different modes in the JAM model. The results computed from different modes are with the similar trends. When the $p_{T}$ coverage is enlarged, the cumulant ratios are suppressed with respected to unity, the Poisson expectations. When the $p_{T}$ range is small, the fluctuations are dominated by Poisson statistics and the cumulant ratios are close to the unity.  We also noticed the recent study for baryon number fluctuations due to mean field effects. This was done with a Relativistic Mean Field (RMF) approach~\cite{Kenji_vector}. Both scaler field ($\sigma$), vector field ($\omega$) as well as the liquid-gas phase transition are included in the RMF calculations for a static nuclear medium. As a result, it is found that the $\kappa \sigma^{2}$ is suppressed by the vector field ($\omega$) while an enhancement, due to the liquid-gas phase transition, was identified at a low temperature $T$=21 MeV and large baryon chemical potential $\mu_{B}$= 906 MeV. On the other hand, here we employed a full dynamic approach with a momentum dependent scaler mean field. Relative to the cascade mode, we do not observe any further suppression from the scaler mean field. In the future, it would be interesting to include the vector field in our dynamic calculations for baryon number fluctuations.

\section{Summary}
With the JAM model, we studied the rapidity and transverse momentum dependence for the cumulants and cumulants ratios of net-proton (baryon) distributions in 0-5\% most central Au+Au collisions at $\sqrt{s_{\text{NN}}} = 5 \,\text{GeV}$. The model simulation is performed with three different modes, which are cascade, mean field and attractive scattering orbit, respectively. The results from three different modes show similar suppression trends, when enlarging the rapidity and/or transverse momentum coverage. Those suppressions in net-proton(baryon) cumulant ratios are expected for the conservation of baryon numbers in strong interactions. Therefore, it indicates that the significant increasing of the $\kappa \sigma^{2}$ of net-proton distribution above unity observed in Au+Au collisions at $\sqrt{s_{NN}}=7.7$ GeV measured by the STAR experiment, cannot be explained by the JAM model with effects of the momentum dependent scaler mean field potential or softening of EoS. This study can provide us important information about the non-critical contribution to the fluctuations of net-proton (baryon) and baseline for the QCD critical point search in the large net-baryon density region.

\section*{Acknowledgments}
The work was supported in part by the MoST of China 973-Project No.2015CB856901, NSFC under grant No. 11575069, 11221504.

\bibliography{root}
\bibliographystyle{unsrt}
\end{document}